\documentclass[12pt]{article}
\usepackage{amssymb,amsthm}
\usepackage[cmex10]{amsmath}
\usepackage[longnamesfirst]{natbib}
\usepackage[dvips]{graphicx}
\usepackage{enumerate}
\usepackage{fullpage}
\usepackage{pdfsync}
\usepackage{colordvi}
\allowdisplaybreaks
\usepackage[nodisplayskipstretch]{setspace}
\usepackage{microtype}
\usepackage{float}
\usepackage[caption = false]{subfig}
\usepackage{multirow}

\usepackage[T1]{fontenc}

\usepackage{lmodern}

\numberwithin{equation}{section}

\usepackage{accents}

\usepackage{verbatim,color}
\definecolor{thistle}{rgb}{0.8,0.05,1}

\def\boxit#1{\vbox{\hrule\hbox{\vrule\kern6pt
          \vbox{\kern6pt#1\kern6pt}\kern6pt\vrule}\hrule}}

\title{Semiparametric Distribution Regression with Instruments and Monotonicity}

\author{Dominik Wied\mythanks{This Version: \today. Institute for Econometrics and Statistics, University of Cologne, e-mail: {\ttfamily dwied@uni-koeln.de.} I am grateful to Alexander Mayer for helpful comments.} \\[1mm]  {\it University of Cologne}}

\date{}

\usepackage{titlesec}
\titleformat{\section}[block]{\centering\normalfont\sffamily}{\thesection.}{0.5em}{\lsstyle\uppercase}
\titleformat{\subsection}[block]{\normalfont\sffamily}{\thesubsection.}{0.4em plus .1em minus .2em}{}
\titleformat{\subsubsection}[runin]{\normalfont\sffamily}{\thesubsubsection.}{0.4em plus .1em minus .2em}{}[.]
\titlespacing*\section{0pt}{18pt plus 4pt minus 2pt}{4pt plus 1pt minus 1pt}
\titlespacing*\subsection{0pt}{16pt plus 3pt minus 2pt}{4pt plus 1pt minus 1pt}
\titlespacing*\subsubsection{0pt}{12pt plus 2pt minus 1pt}{4pt plus 1pt minus 1pt}

\makeatletter
\def\mythanks#1{%
    \protected@xdef \@thanks {\@thanks \protect \footnotetext [\the \c@footnote ]{#1}}%
}
\makeatother

\begin{document}

\newtheorem{theorem}{Theorem}
\newtheorem{definition}{Definition}
\newtheorem{lemma}{Lemma}
\newtheorem{assumption}{Assumption}
\theoremstyle{definition}
\newtheorem{example}{Example}
\newtheorem{remark}{Remark}
\parindent 0cm

\bibliographystyle{ecta}

{
\maketitle

\begin{abstract} \noindent This paper proposes IV-based estimators for the semiparametric distribution regression model in the presence of an endogenous regressor, which are based on an extension of IV probit estimators. We discuss the causal interpretation of the estimators and two methods (monotone rearrangement and isotonic regression) to ensure a monotonically increasing distribution function. Asymptotic properties and simulation evidence are provided. An application to wage equations reveals statistically significant and heterogeneous differences to the inconsistent OLS-based estimator.
\end{abstract}

\noindent \textbf{JEL Classification:} C26, J30

\noindent \textbf{Keywords:} Control function, endogeneity, isotonic regression, wage equations
}
\newpage

\doublespacing

\section{Introduction}
The semiparametric distribution regression (DR) model introduced by \citet{foresiperacchi:1995} has become a popular model for conditional distributions if other quantities than only the conditional expectation are of interest. An important feature of this model is that no distribution assumptions on the response are made, e.g. $Y$ is not assumed to be normally distributed, conditionally on covariates. At the same time, the model provides interpretable functional forms between the regressors and the outcomes, while estimating the conditional response distribution semi-parametrically. From the estimated distribution function, quantiles could be directly obtained by inversion.

One typical application are conditional wage distributions, where upper or lower quantiles are supposed to be modelled. \citet{chernozhukov:2013} and \citet{rothewied:2013} show that the DR model might be better suited than quantile regression for handling certain characteristics of wage data such as genuine point masses in the distribution of wages, nonlinearities around the minimum wage and rounding effects. The appealing property is that e.g. censoring points do not have to be included ex ante as in the case of censored quantile regression, but are detected by the estimation itself. \citet{chernozhukov:2013} show how the model can be used for estimating counterfactual distributions, \citet{rothewied:2020} propose a method for estimating conditional densities and quantile partial effects in this model. See also \citet{koenker:2013} for a comparison of quantile and distribution regression.

A restriction of the literature up to now is that the regressors are assumed to be exogenous. For example, \citet{rothewied:2013} consider a version of Mincer's earnings function by explaining the logarithmic wage with the years of education and the years of experience among others, not taking into account that, for example, the years of education might be an endogenous regressor. This does not mean that the DR estimates in such approaches are not useful. They do estimate conditional distribution functions consistently, but there is no control for unobserved confounders. A particular value of the years of education is correlated to some degree of ability or motivation of employee, so that one gets the distribution only for a subset of the population. The novelty of the present approach is the control for confounders, so that we get a clearer picture of the population.

There are some recent papers on DR estimation with endogenous regressors. \citet{sanchez:2020} consider DR estimation based on instrumental variables, but they use parametric models based on splines among others. \citet{chernozhukov:2022} discuss semiparametric DR models in the context of sample selection.

The present paper proposes IV-based estimators for the semiparametric DR model. Taking into account that the DR model is fitted by pointwise estimators of simple binary outcome models, we adapt consistent estimators for binary outcome models with endogenous regressors. On the one hand, we consider maximum likelihood estimation, which is asymptotically efficient, on the other hand, we propose a computationally better tractable three-step estimator. For both estimators, consistency and convergence to Gaussian limit processes are proved. As these estimator are unconstrained, monotonicity is not guaranteed. We discuss two methods for enforcing monotonicity in a second step, monotone rearrangement and isotonic regression.

In the following, we first present the model (Section 2), then the estimation procedures including a causal interpretation and asymptotic results (Section 3). Afterwards. we consider the monotonizing methods (Section 4) and some simulation evidence (Section 5). An application to a Mincer-type wage regression (Section 5) demonstrates the importance in empirical practice to take endogeneity into account for estimating DR models and to use the new method. Section 6 makes some suggestions for future research.

\section{Model}
Consider an outcome variable $Y$ and regressors $X_1,\ldots,X_k$. In the semiparametric DR model, the conditional distribution function of $Y$ given the set of regressors $X$ is modelled by $F_{Y|X}(y|x) = \Lambda(x'\beta(y))$ for some link function $\Lambda$ such as the distribution function of the standard normal distribution $\Phi$ and some function $\beta(y)$. Although the function $\Lambda$ must be chosen in advance, in this note, the model is called semiparametric: Usually, in the literature, no explicit restrictions on $\beta(y)$ such as continuity are imposed and there is a parameter for every $y$. Anyway, it is clear that $\beta(y)$ has to be chosen such that $F_{Y|X}(y|x)$ fulfills the properties of a conditional distribution function if the model is correctly specified. Given $x$, the function must be monotonically increasing and converge to $1$ and $0$ for $y \rightarrow \infty$ and $y \rightarrow -\infty$, respectively.

In the simple linear model $Y = 1 + X + U$, where $U$ is distributed with distribution function $\Lambda$ and independent of $X$, it holds $F_{Y|X}(y|x) = \Lambda(y-1-x)$ such that $\beta(y) = (y-1,-1)$ and the monotonicity condition is fulfilled. If, for example, $Y$ describes earnings of employees and there is minimum wage at some value $y^*$, the function $\beta(y)$ might contain a discontinuity point at $y^*$.

Based on an i.i.d. sample of length $n$, $F_{Y|X}(y|x)$ can be consistently estimated by maximum likelihood estimation similarly as a probit model would be estimated, for example. The estimation is performed separately for each $y$ and requires that the regressors are exogenous. To be precise, one introduces the indicator functions $I_y := 1\{Y \leq y\}$ with $E(I_y|X=x) = \Lambda(x'\beta(y))$. The model can be interpreted as a latent variable model with $I_y = 1$ if $I_y^* := X'\beta(y) \geq U$ and $I_y = 0$ otherwise. The random variable $U$ is distributed with distribution function $\Lambda$ and exogeneity means that $X$ is independent of $U$.

\section{Estimation Procedure}

There are different IV-based approaches for estimating binary outcome models if $X$ and $U$ are not independent. We present two adaptions for the case of estimating such models separately for each $y$. The first one is based on the maximum likelihood estimator, which is asymptotically efficient for fixed $y$. As this estimator is computationally demanding, we then propose a three-step estimator. This is an adaption of the two-step estimator introduced by \citet{riversvuong:1988}, which, for fixed $y$, is also asymptotically efficient in the case of just identified models.

\subsection{Maximum Likelihood Estimator}

The original maximum likelihood estimator was brought forward and disussed by \citet{amemiya:1978}, \citet{newey:1987}, \citet{riversvuong:1988}, is explained in detail in \citet{wooldridge:2002}, Section 15.7.2 and \citet{hansen:2022}, Section 25.12., and is implemented in Stata (command \textit{ivprobit}). The estimator is applicable if $\Lambda$ is equal to the distribution function of a standard normal distribution $\Phi$ and if the endogenous regressors are continuously distributed. We focus on the case of one endogenous regressor. Using the notation from the literature, $X$ denotes a $k$-dimensional vector with exogenous regressors, $Y_2$ is scalar, endogenous and continuously distributed and $Z$ is a $l$-dimensional vector with exogenous instruments.

In our situation, the goal is to estimate $E_V(I_y|X=x,Y_2=y_2) := P_V(Y \leq y| X=x,Y_2=y_2) := E_V(I_y|X=x,Y_2=y_2,V=v)$, i.e. we first control for possible confounders (such as ability/motivation) and integrate these confounders out afterwards. Roughly spoken, $Y_2$ is made exogenous by this integration. Standard probit would be a suitable estimator for the conditional distribution function $E(I_y|X=x,Y_2=y_2)$.

The assumption is that $E_V(I_y|X=x,Y_2=y_2) = P_V(I_y^* \geq 0|X=x,Y_2=y_2)$ with
\begin{eqnarray}\label{model}
I_y^* &=& X'\beta_1(y) + Y_2 \beta_2(y) + U(y) \\\nonumber
Y_2 &=& X'\gamma_1 + Z'\gamma_2 + V,
\end{eqnarray}
Here, the random variables $U(y)$ and $V$ are jointly normally distributed conditionally on $X$ and $Z$,

\begin{equation*}
\begin{pmatrix} U(y) \\ V \end{pmatrix} | (X,Z) \sim \mathcal{N} \left(\begin{pmatrix} 0 \\ 0 \end{pmatrix},\begin{pmatrix} 1 & \sigma_{12} \\ \sigma_{12} & \sigma_2^2 \end{pmatrix} \right).
\end{equation*}

Note that the variance of $U$ is set to $1$ because an additional parameter would not be identified. With the explanations above, $P_V(Y \leq y | X=x,Y_2=y_2) = \Phi(x'\beta_1(y) + y_2 \beta_2(y))$.

\begin{remark}
To get more intuition on this, note that one can write $U(y) = \rho V + \epsilon(y)$, where $\epsilon \sim \mathcal{N}(0,\sigma_\epsilon^2)$ is independent from $V$, $\rho := \frac{\sigma_{12}}{\sigma_2^2}$ and $\sigma_\epsilon^2 := 1-\rho^2\sigma_2^2$. Then we have $I_y^* = X'\beta_1(y) + Y_2 \beta_2(y) + \rho V + \epsilon(v)$ and $E(I_y^*|X=x,Y_2=y_2,V) = x'\beta_1(y) + y_2 \beta_2(y) + \rho V$- The expectation of the latter term with respect to $V$ is $x'\beta_1(y) + y_2 \beta_2(y)$ because $E(V)=0$. In contrast to this, assuming joint normality of $(U(y),V,X,Z)$, $U(y) = \psi Y_2 + C(y)$ for $\psi = \frac{\sigma_{12}}{\sigma_{Y_2}^2} = \frac{\rho \sigma_2}{\sigma_{Y_2}^2}$, where $C(y)$ is independent of $Y_2$ and $\sigma_{Y_2}^2 = Var(Y_2)$. Then, similarly as in \citet{li:2022},

\begin{equation*}\label{conddistrfunc}
P(Y \leq y | X=x,Y_2=y_2) = \Phi\left(\frac{x'\beta_1(y) + \left(\beta_2(y) + \psi \right) y_2}{\sqrt{1-\tau^2}}\right)
\end{equation*}
with $\tau = \frac{\sigma_{12}}{\sigma_{Y_2}} = \frac{\rho \sigma_2}{\sigma_{Y_2}}$. This is the quantity that standard probit would estimate, but this is not the quantity we are interested in. \hfill $\Box$
\end{remark}

This means that consistent estimation of the parameters in \eqref{model} leads to consistent estimation of $E_V(I_y|X=x,Y_2=y_2)$ by the continuous mapping theorem.

For estimation purposes, we use the i.i.d. sample $(I_{y,i},Y_{2,i},X_i,Z_i)$ with $I_{y,i} = 1\{Y_i \leq y\}$. Similarly as in \citet{hansen:2022}, the likelihood is derived by factorizing the joint density of $I_y$ and $Y_2$. The log-likelihood is then essentially the sum of the standard regression and the standard probit log-likelihood. It is given as $L_y(\theta(y)) = \sum_{i=1}^n L_{y,i}(\theta(y))$ with the parameter vector\footnote{Note that $\sigma_{12}$ and $\sigma_\epsilon^2$ can be calculated from the other parameters.} $\theta(y) := (\beta_1(y),\beta_2(y),\gamma_1,\gamma_2,\rho,\sigma_2^2)$ and
\begin{eqnarray*}
L_{y,i}(\theta(y)) &=& I_{y,i} \log \Phi\left(\frac{\mu_{y,i}(\theta(y))}{\sigma_\epsilon} \right) + (1-I_{y,i}) \log \Phi\left(1-\frac{\mu_{y,i}(\theta(y))}{\sigma_\epsilon} \right) \\
                   && -\frac{1}{2} \log(2 \pi) -\frac{1}{2} \log \sigma_2^2 - \frac{1}{2 \sigma_2^2} \left(Y_{2,i} - X_i'\gamma_1 - Z_i'\gamma_2\right)^2.
\end{eqnarray*}
It holds $\mu_{y,i}(\theta(y)) = X_i'\beta_1(y) + Y_{2,i} \beta_2(y) + \rho(Y_{2,i} - X_i'\gamma_1 - Z_i'\gamma_2 )$ and $\sigma_\epsilon = \sqrt{1-\rho^2\sigma_2^2}$.

For each $y$, the parameter estimator can be equivalently calculated by maximizing the likelihood function or by minimizing some norm of the score function. Thus, the estimator falls into the framework of Z-estimators analyzed in \citet{chernozhukov:2013} and one can derive consistency and asymptotic normality, both pointwisely and uniformly in $y$. The estimator for the conditional distribution function\footnote{For better readability, we do not write $F_V$ in the following, which would be the coherent notation.} $F_{Y|X,Y_2}(y|x,y_2) := \Phi(x'\beta_1(y) + y_2 \beta_2(y))$ is given by $\hat F_{ML,Y|X,Y_2}(y|x,y_2) := \Phi(x'\hat \beta_1(y) + y_2 \hat \beta_2(y))$. This is a continuous transformation and the limit results carry over by means of the functional delta method. Then, under some additional assumptions as described in the Appendix, we obtain

\begin{theorem}\label{theorem1}
Let Assumption 1 be fulfilled. Then it holds that
\begin{equation}\label{eq:theorem1}
\sqrt{n}\left(\hat F_{ML,Y|X,Y_2}(\cdot|x,y_2)- F_{Y|X,Y_2}(\cdot|x,y_2) \right)
\end{equation}
converges to a Gaussian process $\mathbb{G}_1(\cdot)$ on a compact subinterval of $\mathbb{R}$.
\end{theorem}
The proof of this theorem shows that the limit process depends both on the limit properties of $\hat \theta(y)$ and the shape of the function $F_{Y|X,Y_2}(y|x,y_2)$.

\subsection{Three Step Approach}
The two step estimator by \citet{riversvuong:1988} is explained in detail in \citet{wooldridge:2002}, Section 15.7.2. The estimator requires a non-trivial adjustment to our situation, however, because it does not directly estimate the parameters $\beta_1(y)$ and $\beta_2(y)$ consistently. The numerical calculations later on are performed with this estimator because it works much faster and the problem of boundary solutions does not appear. The maximum likelihood estimator might be used in special situations, in which one prefers to estimate the conditional distribution function only for a single point $y$, say.

The setup is similar to the one in the former subsection and the estimator is based on the decomposition $U(y) = \rho V + \epsilon(y)$ which leads to the equation $$I_y^* = X'\beta_1(y) + Y_2 \beta_2(y) + V \frac{\rho}{\sigma_2} + \epsilon(y).$$ The error term $\epsilon(v)$ is independent of $X$, $Y_2$ and $V$ and is $N(0,1-\rho^2)$-distributed. This means that, for fixed $y$, standard probit estimation would consistently estimate the parameters $\tilde{\beta}_1(y):= \frac{\beta_1(y)}{\sqrt{1-\rho^2}}, \tilde{\beta}_2(y) := \frac{\beta_2(y)}{\sqrt{1-\rho^2}}, \tilde{\rho} := \frac{\rho}{\sigma_2\sqrt{1-\rho^2}}$ under the same assumptions as discussed after Assumption 1 in the Appendix. As $V$ is not observable, this term is replaced with the residuals of an OLS regression of $Y_2$ on $X$ and $Z$. 

For fixed $y$, this is the two step estimator by \citet{riversvuong:1988}. In the case of just identified models (one instrument for the endogenous regressor), this estimated is even numerically equal to the maximum likelihood estimator for $\tilde{\beta}_1(y),\tilde{\beta}_2(y),\tilde{\rho}$, so that Theorem \ref{theorem1} can be directly applied to this.\footnote{This is also true for the AGLS estimator from \citet{amemiya:1978}, which is implemented in the \textit{R}-package \textit{ivprobit}.}

The parameter $\rho$ is not known, so that these estimators cannot be used directly. However, they can be used to consistently estimate the conditional expectation $$E(I_y|X=x,Y_2=y_2,V=v) = \Phi(x'\tilde{\beta}_1(y) + y_2 \tilde{\beta}_2(y) + v \tilde{\rho}).$$ As $E(I_y|X=x,Y_2=y_2) = E_V(I_y|X=x,Y_2=y_2,V)$ by the law of iterated expectations, a consistent estimator for $E'(I_y|X=x,Y_2=y_2)$ is given by $$\hat F_{Y|X,Y_2}(y|x,y_2) := \frac{1}{n} \sum_{i=1}^n \Phi(x'\widehat{\tilde{\beta}_1(y)} + y_2 \widehat{\tilde{\beta}_2(y)} + V_i \widehat{\tilde{\rho}}),$$ where $V_i$ are the residuals of an OLS regression of $Y_{2i}$ on $X_i$ and $Z_i$, $i=1,\ldots,n$.

\begin{theorem}\label{theorem2}
Let Assumption 1 be fulfilled and consider the case of a just identified model. Then it holds that
\begin{equation}\label{eq:theorem2}
\sqrt{n}\left(\hat F_{Y|X,Y_2}(\cdot|x,y_2)- F_{Y|X,Y_2}(\cdot|x,y_2) \right)
\end{equation}
converges to a Gaussian process $\mathbb{G}_2(\cdot)$ on a compact subinterval of $\mathbb{R}$.
\end{theorem}

Due to the discretization in the estimation, the estimator $\hat F_{Y|X,Y_2}(y|x,y_2)$ can attain at most $n$ different values for fixed $X$ and $Y_2$. The differences arise at the different outcomes $Y_i$, so that it is reasonable to evaluate the estimated distribution function at all $Y_i$, if computationally feasible.

\section{Monotonicity}
While the proposed estimators from the last section are consistent under appropriate assumptions, there is no reason to assume that the estimated conditional distribution functions are monotonically increasing in $y$ in finite samples. This might be a drawback for interpretation purposes, e.g. if the estimators are used for calculating conditional quantiles and it turns out that the estimated $90\%$-quantile is smaller than the estimated $80\%$-quantile. We discuss two methods to fix this, monotone rearrangement as well as isotonic regression.\footnote{\citet{foresiperacchi:1995} discuss in their Section 2.1 some other possibilities to get monotonous estimators of the distribution function, but do not elaborate on them in more detail.} While the former has well-known asymptotic properties, the latter is computationally more appealing. A simulation study reveals that both approaches share similar properties in terms of the mean squared error.

\subsection{Monotone Rearrangement}
\citet{chernozhukov:2010} propose a monotone rearrangement approach, mainly for quantile regression in order to ensure that estimated conditional quantiles do not cross. As discussed in \citet{chernozhukov:2013}, this approach can also be applied to distributional regression. It is based on the identity
\begin{equation}\label{MonotoneOperator}
F_{Y|X,Y_2}(y|x,y_2) = \int_0^1 \mathbf{1}\{Q_{Y|X,Y_2}(u|x,y_2) \leq y\}du,
\end{equation}
so that in a first step the conditional quantile function needs to be estimated, before it is appropriately integrated. This leads to the estimator $$\tilde F_{Y|X,Y_2}(y|x,y_2) = \int_0^1 \mathbf{1}\{\hat Q_{Y|X,Y_2}(u|x,y_2) \leq y\}du$$ with the estimated conditional quantile function\footnote{A researcher only interested in conditional quantiles could of course directly use this estimator.} $\hat Q_{Y|X,Y_2}(u|x,y_2) = \mathsf{inf}_y \{\hat F_{Y|X,Y_2}(y|x,y_2) \geq u \}$. The asymptotic properties of this estimator are well understood. As discussed in \citet{chernozhukov:2010}, given a result like Theorem \ref{eq:theorem1} from the last section, the convergence rate (in our case $\sqrt{n}$) carries over due to the Hadamard differentiability of the operator from \eqref{MonotoneOperator} and an application of the functional delta method. Moreover, it is possible to estimate the limit process by a bootstrap approximation.

\subsection{Isotonic Regression}

An alternative to the monotone rearrangement is the application of an isotonic regression, which can be applied directly on the functional estimator. This estimation procedure is discussed in \citet{barlow:1972} and \citet{robertson:1988}, for example. By construction, the estimated distribution function only changes its value at the observed $Y_1,\ldots,Y_n$ and is constant between these points. The idea is to replace the points $\hat F_i := \Phi(x'\hat \beta_1(Y_i) + y_2 \hat \beta_2(Y_i))$ by points $\tilde{\tilde{F_i}}$ that are close to $\hat F_i$, but fulfill the monotonicity restriction. This means that one solves the quadratic minimization problem

\begin{equation*}
min_{\tilde{\tilde{F_1}},\ldots,\tilde{\tilde{F_n}}} \sum_{i=1}^n \left(\tilde{\tilde{F_i}} - \hat F_i \right)^2
\end{equation*}
under the constraint

\begin{equation}\label{constraint}
\tilde{\tilde{F_i}} \leq \tilde{\tilde{F_j}} \text{ for } Y_i \leq Y_j.
\end{equation}

The problem can be solved numerically with the {\it pool adjacent violators algorithm}, an implementation in software packages such as R (command \textit{isoreg} in the package \textit{stats}) is available. The computational complexity for given $n$ is $O(n)$ for already sorted data, see \citet{best:1990}. So, the approach is less complex than the monotone arrangement, where the quantile function has to be used and where integrals have to be solved. A potential drawback is the tendency to obtain flat functions, which leads to a bias in finite samples, if the true distribution function is strictly increasing.

Having obtained a monotonically increasing distribution function for the points $Y_1,\ldots,Y_n$, forecasts for other values of $y$ might be obtained by linear interpolation, for example. Also conditional quantiles can be calculated in this way.

If \eqref{constraint} already holds for the $\hat F_i, i=1,\ldots,n,$ $\sum_{i=1}^n \left(\tilde{\tilde{F_i}} - \hat F_i \right)^2$ is equal to $0$. So, it is intuitive that the monotonized estimator is consistent if the true conditional distribution function is monotonically increasing and the estimated distribution function is uniformly consistent (over $y$).

In other contexts, the convergence rate of isotonic regression is smaller than $\sqrt{n}$, for example $n^{1/3}$ in \citet{abrevaya:2005}. In these cases, the standard bootstrap (drawing with replacement) might behave erratic, see \citet{patra:2018}. Monte Carlo evidence in the following subsection suggests that such 5problems should not expected the present context, at least not for the setting considered in the empirical application. The intuition is that in our case, the isotonic regression is just a finite-sample correction in second step of an estimator which asymptotically fulfills the monotonicity restriction.

\section{Simulations}
We simulate from the model $Y^* = \mathsf{max}(2,\tilde Y)$ and
\begin{eqnarray*}
\tilde Y &=& 1+X+Y_2+U \\
Y_2 &=& 1+X+Z+V,
\end{eqnarray*}
where $X$ and $Z$ are i.i.d. $N(0,1)$-distributed and $(U,V)$ is bivariate normally distributed with zero mean and covariance matrix $\begin{pmatrix} 1 & \rho \\ \rho & 1 \end{pmatrix}$. Here, $X$ represents the exogenous regressor, $Y_2$ the endogenous regressor, which is correlated with $U$, and $Z$ the exogenous instrument. We consider a censored $\tilde Y$, which mimics the application of modelling wages with a minimum wage and which shall highlight the appealing property of distributional regression of detecting such censoring points. In this case, $F_{Y|X,Y_2}(y|x,y_2) = \Phi(y-1-x-y_2)$ for $y \geq 2$ and $0$ elsewhere. We fix $\rho=0.7$ and calculate $\tilde F_{Y|X,Y_2}(y|x,y_2)$ (monotone rearrangement) and $\tilde{\tilde{F}}_{Y|X,Y_2}(y|x,y_2)$ (isotonic regression) for $x=y_2=1$ and $x=y_2=2$. As $E(X)=0$ and $E(Y_2)=1$, $X$ and $Y_2$ are further away from their expectations in the latter case. The grid points for $y$ are equidistant in the interval $[1,5]$ with $50$ grid points in total. For the rearrangement, the quantile levels are equidistant in the interval $[0.01,0.99]$ with $99$ grid points in total. To mimic the setting of the empirical application, the sample sizes are $n=100,200,400$. For each case, $1000$ Monte Carlo replications are performed. The results are compared with the standard probit estimates that ignore the endogeneity.

As we are concerned with uniform convergence to the true function (see Theorem \ref{theorem1}), we consider the average squared bias, the average variance and the average MSE of $\tilde F_{Y|X,Y_2}(y|x,y_2)$ and $\tilde{\tilde{F}}_{Y|X,Y_2}(y|x,y_2)$ over the grid of $41$ y-values. Tables \ref{table:sim} shows the results.

\begin{center}
- Table \ref{table:sim} here -
\end{center}

\begin{table}[!h!]
\begin{center}
\begin{tabular}{c|c|c|c|c|c|c|c}
Values for $x=y_2$ & $n$ & \multicolumn{3}{c}{Monotone rearrangement} & \multicolumn{3}{c}{Isotonic regression} \\\hline
& & Bias$^2$ & Var & MSE & Bias$^2$ & Var & MSE \\\hline
\multicolumn{8}{c}{OLS} \\\hline
  & 100 & 0.0119 & 0.0085 & 0.0205 & 0.0102 & 0.0083 & 0.0185 \\
1 & 200 & 0.0108 & 0.0036 & 0.0145 & 0.0100 & 0.0035 & 0.0136 \\
  & 400 & 0.0099 & 0.0017 & 0.0116 & 0.0094 & 0.0017 & 0.0116 \\\hline
  & 100 & 0.0076 & 0.0169 & 0.0245 & 0.0043 & 0.0161 & 0.0205 \\
2 & 200 & 0.0049 & 0.0076 & 0.0125 & 0.0038 & 0.0075 & 0.0113 \\
  & 400 & 0.0048 & 0.0039 & 0.0087 & 0.0044 & 0.0039 & 0.0083 \\\hline
\multicolumn{8}{c}{IV} \\\hline
  & 100 & 0.0004 & 0.0098 & 0.0102 & $5 \cdot 10^{-5}$ & 0.0094 & 0.0094 \\
1 & 200 & $8 \cdot 10^{-5}$ & 0.0044 & 0.0045 & $2 \cdot 10^{-5}$ & 0.0042 & 0.0043 \\
  & 400 & $1 \cdot 10^{-5}$ & 0.0022 & 0.0022 & $< 1 \cdot 10^{-5}$ & 0.0021 & 0.0021 \\\hline
  & 100 & 0.0007 & 0.0121 & 0.0128 & $7 \cdot 10^{-5}$ & 0.0103 & 0.0104 \\
2 & 200 & $4 \cdot 10^{-5}$ & 0.0049 & 0.0050 & $< 1 \cdot 10^{-5}$ & 0.0047 & 0.0047 \\
  & 400 & $1 \cdot 10^{-5}$ & 0.0023 & 0.0024 & $< 1 \cdot 10^{-5}$ & 0.0023 & 0.0023 \\\hline
\end{tabular}
\end{center}
\caption{Average squared bias, squared variance and squared MSE of the two monotonizing approaches with OLS probit and IV probit}\label{table:sim}
\end{table}

With the IV approach, the average MSE is dominated by the variance and is similar for both procedures with a slight advantage for the isotonic regression. Bias, variance and MSE are slightly higher if $x$ and $y_2$ are further away from their expectations and halve when the sample size is doubled. This suggests that, in this setup, the convergence rate of both estimators is $\sqrt{n}$. The variance of the OLS approach also halves with doubled sample size and slightly exceeds that of the IV approach for $x=y_2=1$, but is considerably biased as expected. So, its MSE is much higher than that of the IV approach.

\section{Application to Wage Equations}
We revisit wage data from \citet{mroz:1987} with $n=428$ individuals, who were working in 1975, and estimate a Mincer-type regression to estimate the returns of education. To be precise, the logarithmic hourly wage is explained by the years of education and the years of working experience (the latter both linearly and quadratically). The variable years of education is assumed to be endogenous as it might be correlated with unobserved variables such as ability or motivation. While there might be some correlation with the years of working experience as well, we assume that other influences are more relevant in that case, so that we assume this variable to be exogenous.

A possible instrument for the years of education is the years of education of the mother. In this dataset, the first stage $F$-statistic is given by approximately $75$ so that the instrument can be assumed to be sufficiently strong. See \citet{wooldridge:2016} for some discussion why this model might be reasonable.

First, we estimate a simple linear model with OLS and with IV:
\begin{equation*}
log(wage)_i = \beta_0 + \beta_1 educ_i + \beta_2 exper_i + \beta_3 exper^2_i + \varepsilon_i.
\end{equation*}

Table \ref{table:app} shows the estimated coefficients as well as the estimated conditional expectations for the $10\%, 50\%$ and $90\%$ quantiles of \textit{educ} and \textit{exper}, respectively (10 and 4, 12 and 12, 16 and 24). This way, the expected wages are calculated for three groups of employees, the low-educated/low-experienced, the middle-educated/middle-experienced and the high-educated/high-experienced.

\begin{center}
- Table \ref{table:app} here -
\end{center}

\begin{table}[!t!]
\begin{center}
\begin{tabular}{c|c|c|c|c|c|c}
$\hat \beta_0$ & $\hat \beta_1$ & $\hat \beta_2$ & $\hat \beta_3$ & $\hat E(log(wage)|$ &  $\hat E(log(wage)|$ &  $\hat E(log(wage)|$ \\
               &                &                &                & $ed=10,ex=4)$  &  $ed=12,ex=12)$ &  $ed=16,ex=24)$ \\\hline
\multicolumn{7}{c}{OLS}\\
 -0.5220 & 0.1075 & 0.0416 & <-0.0008 & 0.7061 & 1.1498 & 1.7281 \\
 (0.1986) & (0.0141) & (0.0132) & (0.0004) & (0.0664) & (0.0410) & (0.0725) \\\hline
\multicolumn{7}{c}{IV}\\
 0.1982 & 0.0493 & 0.0449 & < -0.0009 & 0.8555 & 1.1948 & 1.5318 \\
 (0.4729) & (0.0374) & (0.0136) & (0.0004) & (0.1115) & (0.0497) & (0.1379) \\
\end{tabular}
\end{center}
\caption{Estimated regression coefficients and conditional expectations for the linear model, standard errors in parentheses}\label{table:app}
\end{table}

Similarly as in other studies with this type of instrument, the IV estimate for \textit{educ} is smaller than the OLS estimate (while the standard error is larger). The intuition is that both the years of education and an unobserved variable which measures ability and/or motivation are positively correlated with the wage, compare also the discussion in \citet{breitung:2022}. The conditional expectations increase if higher values of \textit{educ} and \textit{exper} are considered. Interestingly, the results for OLS and IV are similar for the $50\%$ quantiles. For the $10\%$ quantiles, the IV estimate is larger than the OLS estimate, for the $90\%$ quantile, the IV estimate is smaller. There seems to be a tendency that the variability in terms of the regressor values is lower for the IV estimation. These results will be confirmed and extended by the DR analysis.

\begin{figure}[!h!]
    \centering
    \subfloat[$0.1$-quantile of the regressors]{\includegraphics[width = .5\textwidth]{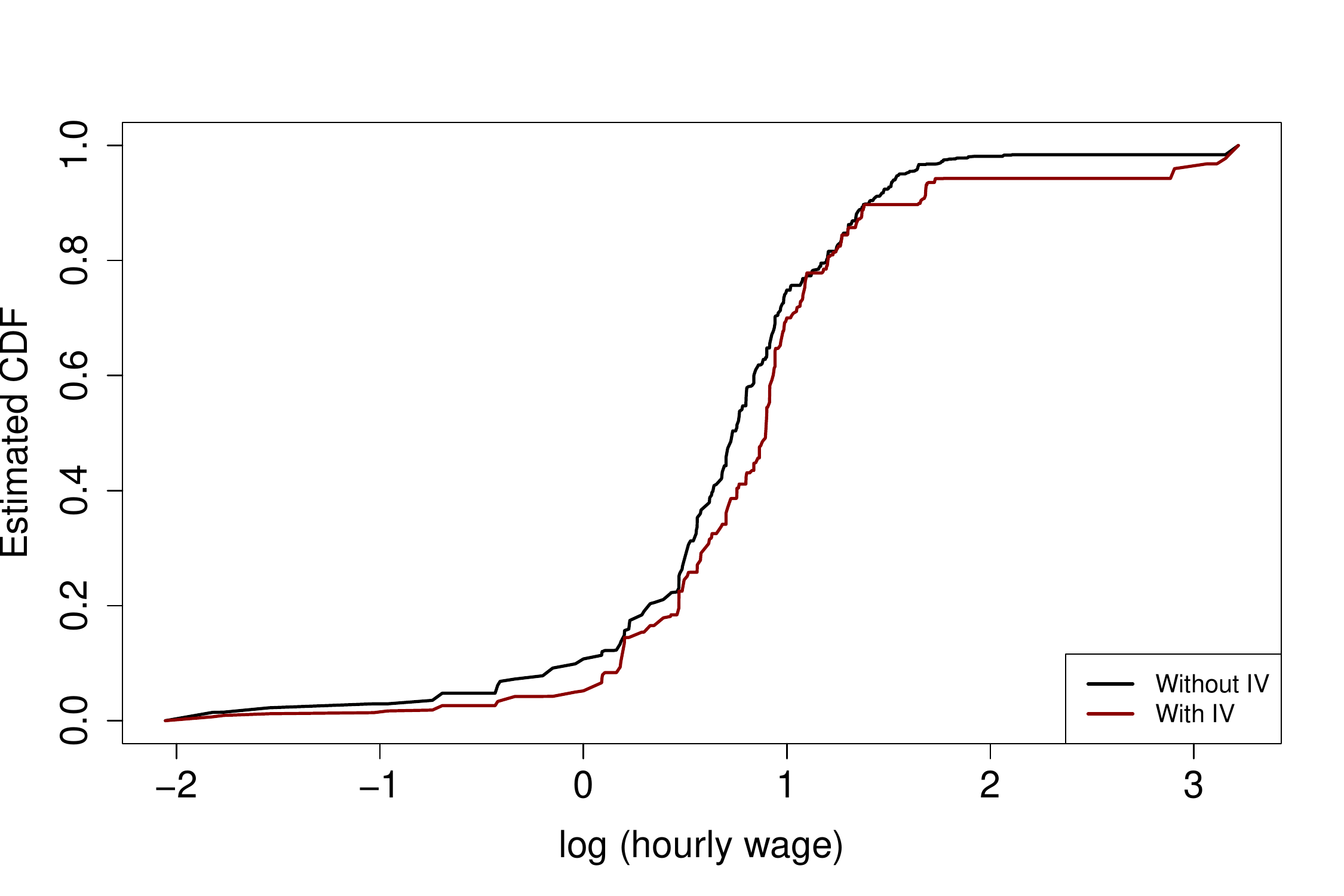}}\\
    \subfloat[$0.5$-quantile of the regressors]{\includegraphics[width = .5\textwidth]{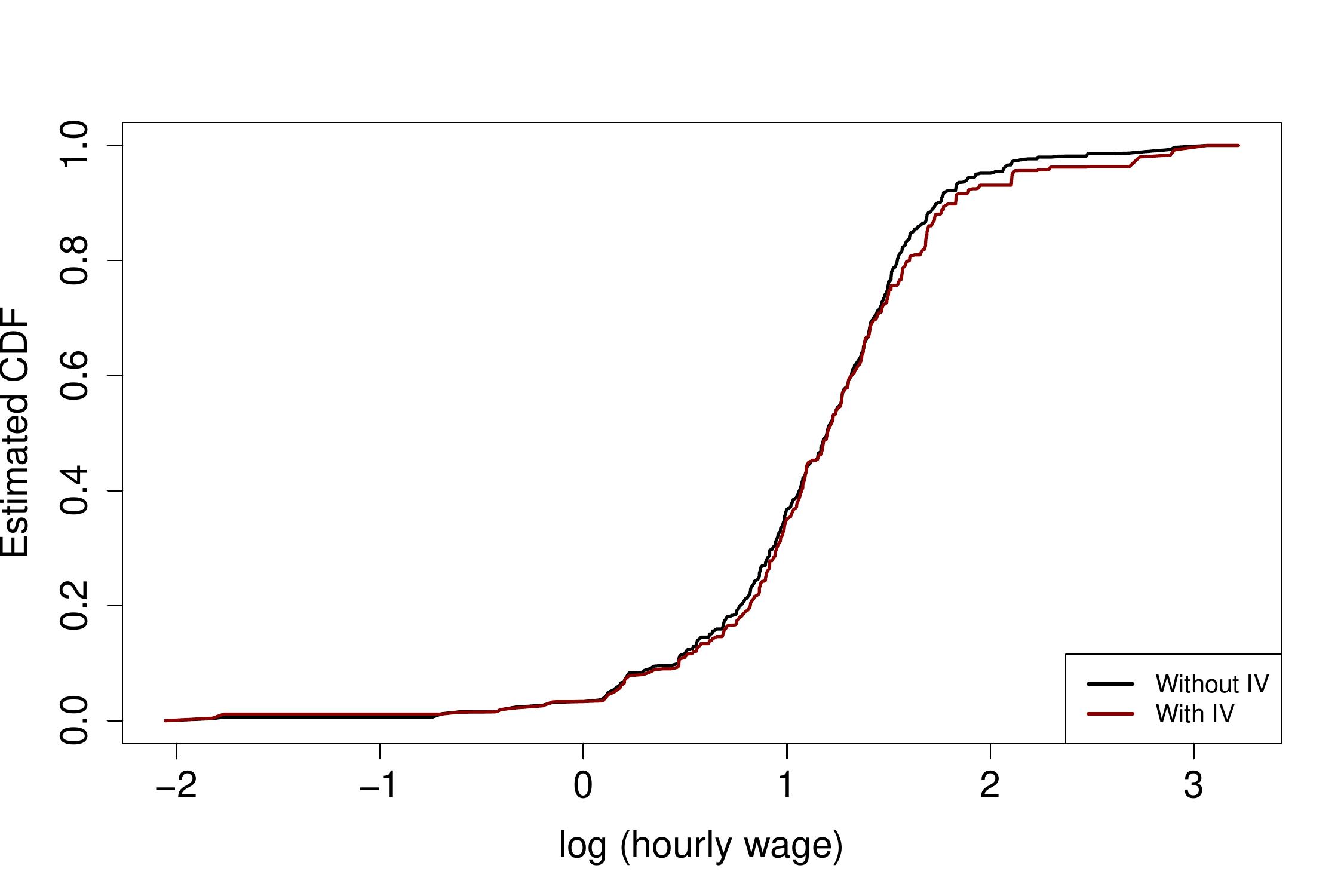}}\\
		\subfloat[$0.9$-quantile of the regressors]{\includegraphics[width = .5\textwidth]{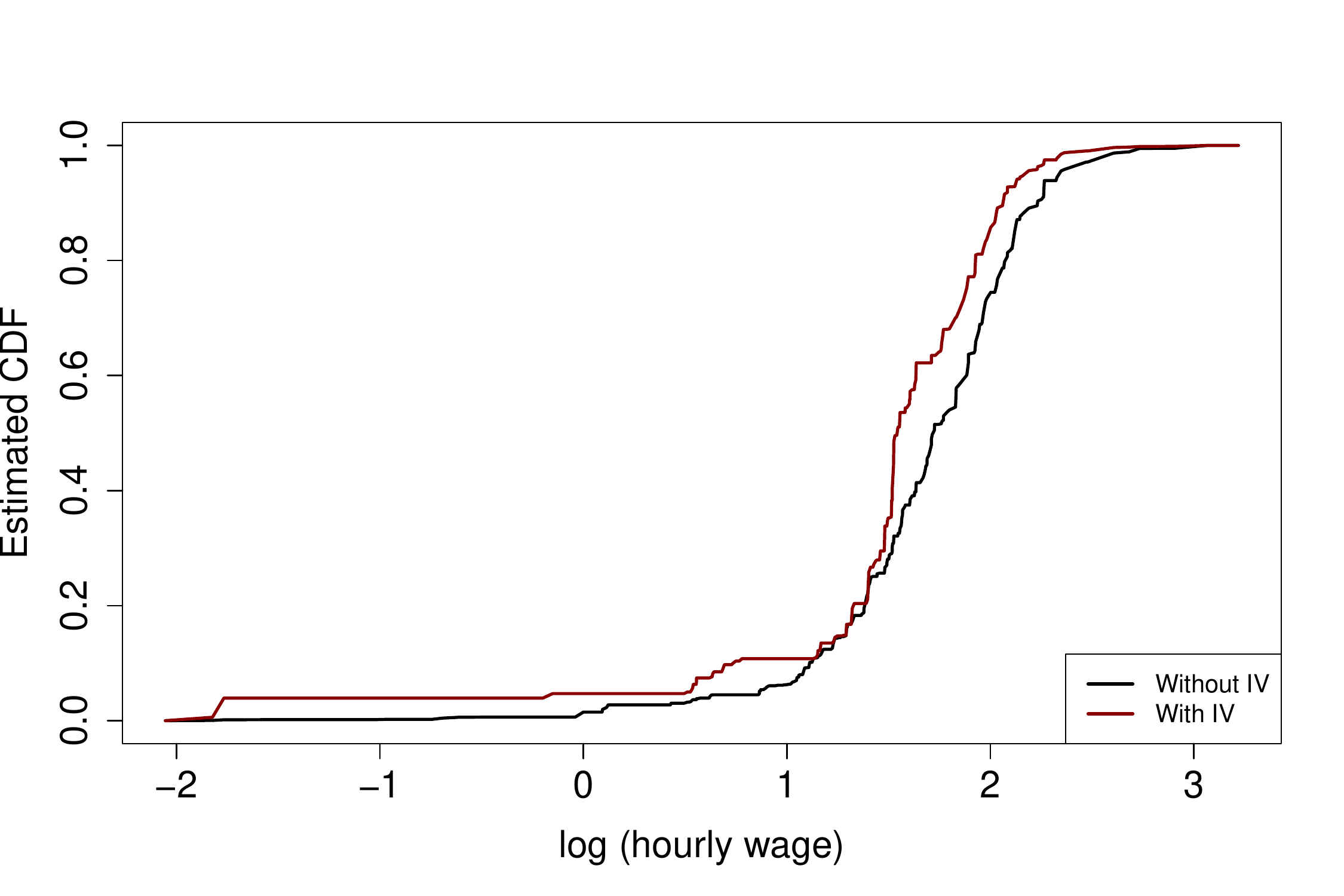}}
    \caption{Estimated conditional distribution functions}\label{figure:app1}
\end{figure}

Figure \ref{figure:app1} shows the estimated conditional distribution functions for both OLS and IV, again for the $10\%, 50\%$ and $90\%$ quantiles of \textit{educ} and \textit{exper}. The estimated distribution functions are evaluated at all outcomes $Y_i$. In all cases, the monotonized version based on isotonic regression discussed in the last section is considered. For higher values of \textit{educ} and \textit{exper}, the distribution functions are shifted more and more to the right. For the $50\%$ quantile, the two functions are rather similar. For the $10\%$ quantile, the IV curve generally lies to the right of the OLS curve, where the largest differences are visible for values of $log(wage)$ between $0.5$ and $1$ as well as around $0$. For the $90\%$ quantile, the IV curve generally lies to the left with the largest differences between $1.5$ and $2$ and the maximal difference is slightly larger than for the $10\%$ quantile.

\begin{center}
- Figure \ref{figure:app1} here -
\end{center}

\begin{figure}[!h!]
    \centering
		\includegraphics[width = .5\textwidth]{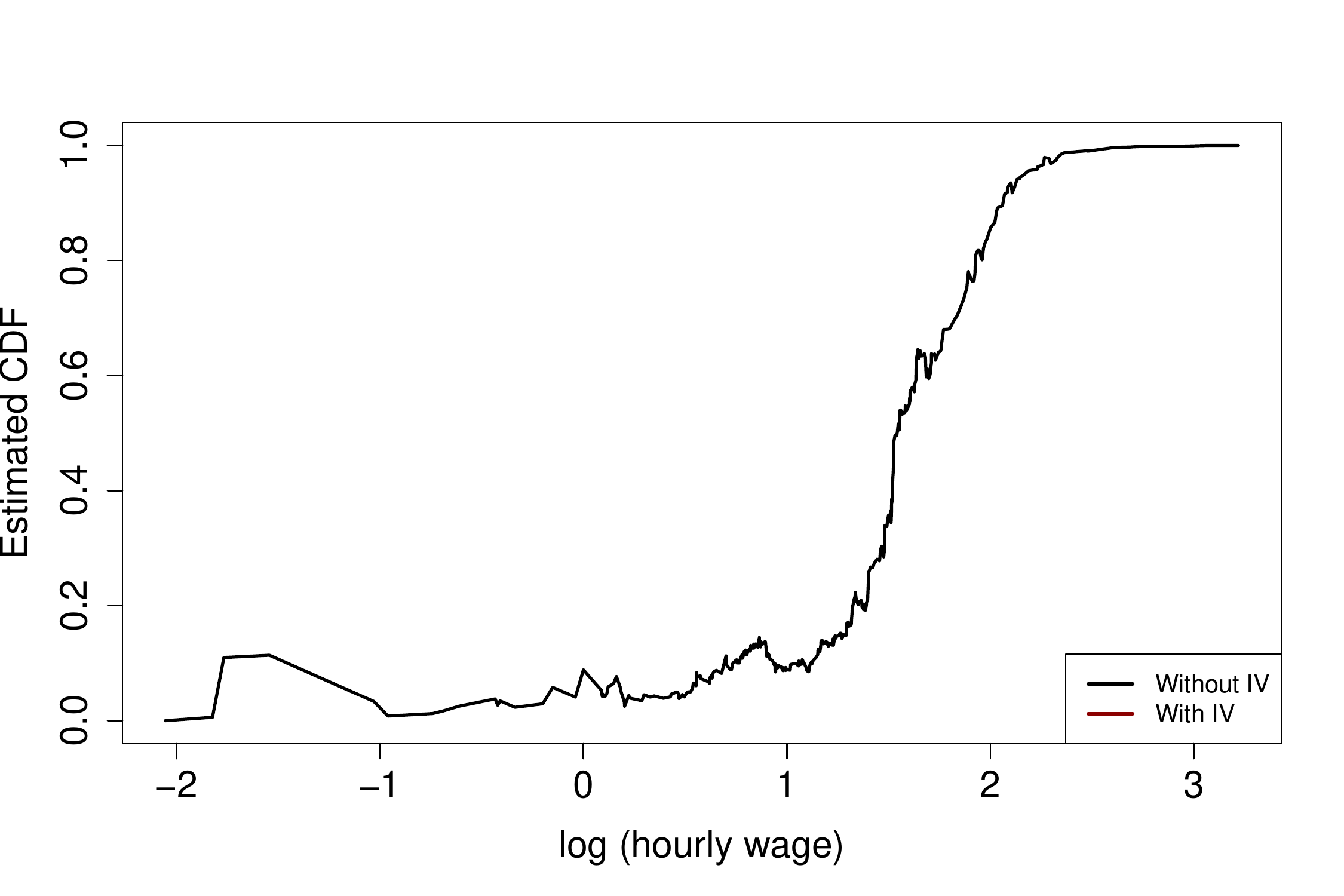}
		\caption{Estimated conditional distribution function without monotonicity constraint for the $0.9$-quantile of the regressors}\label{figure:app1:withoutmono}
\end{figure}

For completeness, Figure \ref{figure:app1:withoutmono} shows the estimated DR curve for the $90\%$ quantiles without monotonization, illustrating why it makes sense to add the monotonizing step.

\begin{center}
- Figure \ref{figure:app1:withoutmono} here -
\end{center}

\begin{figure}[!h!]
    \centering
    \subfloat[$0.1$-quantile of the regressors]{\includegraphics[width = .5\textwidth]{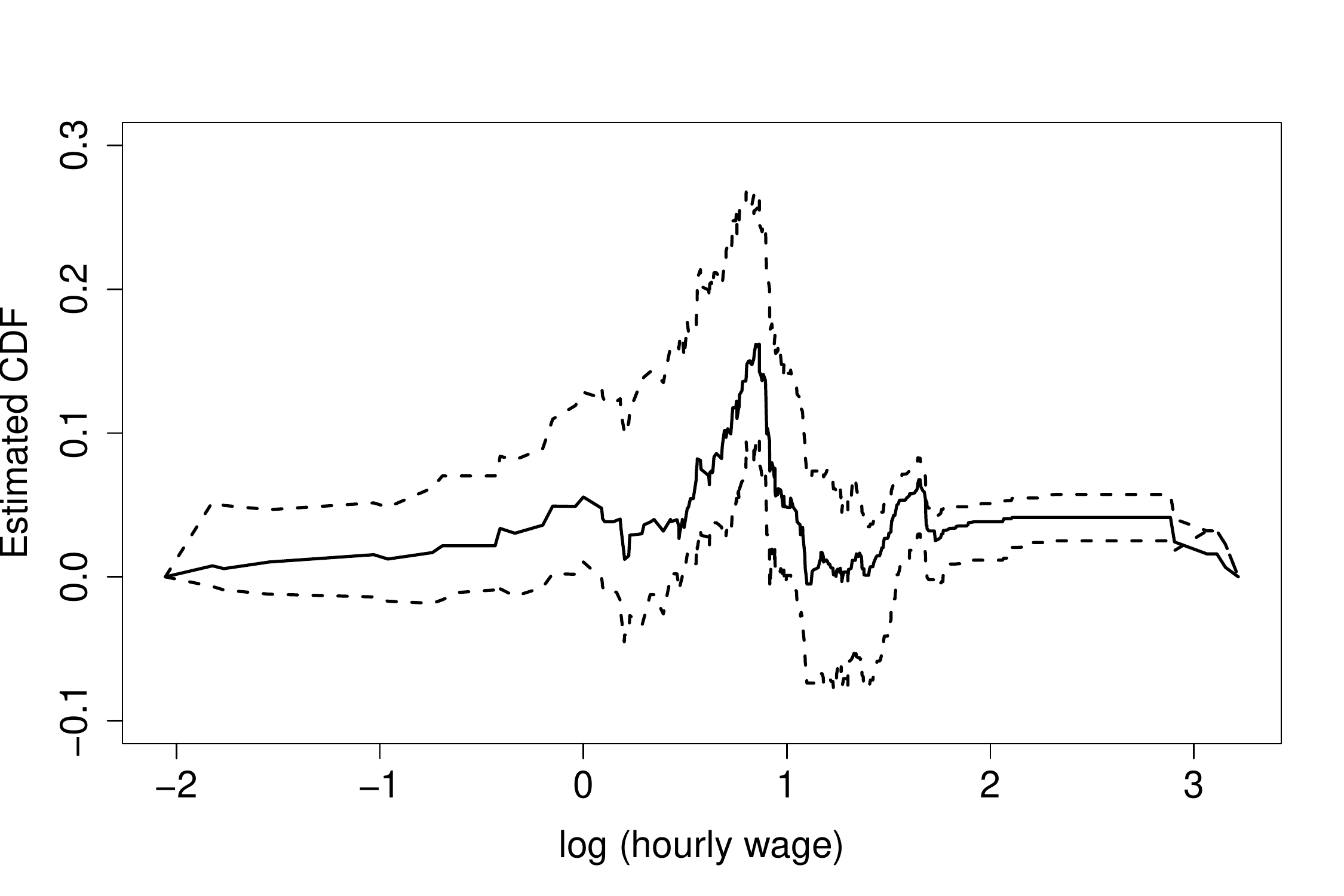}}\\
    \subfloat[$0.5$-quantile of the regressors]{\includegraphics[width = .5\textwidth]{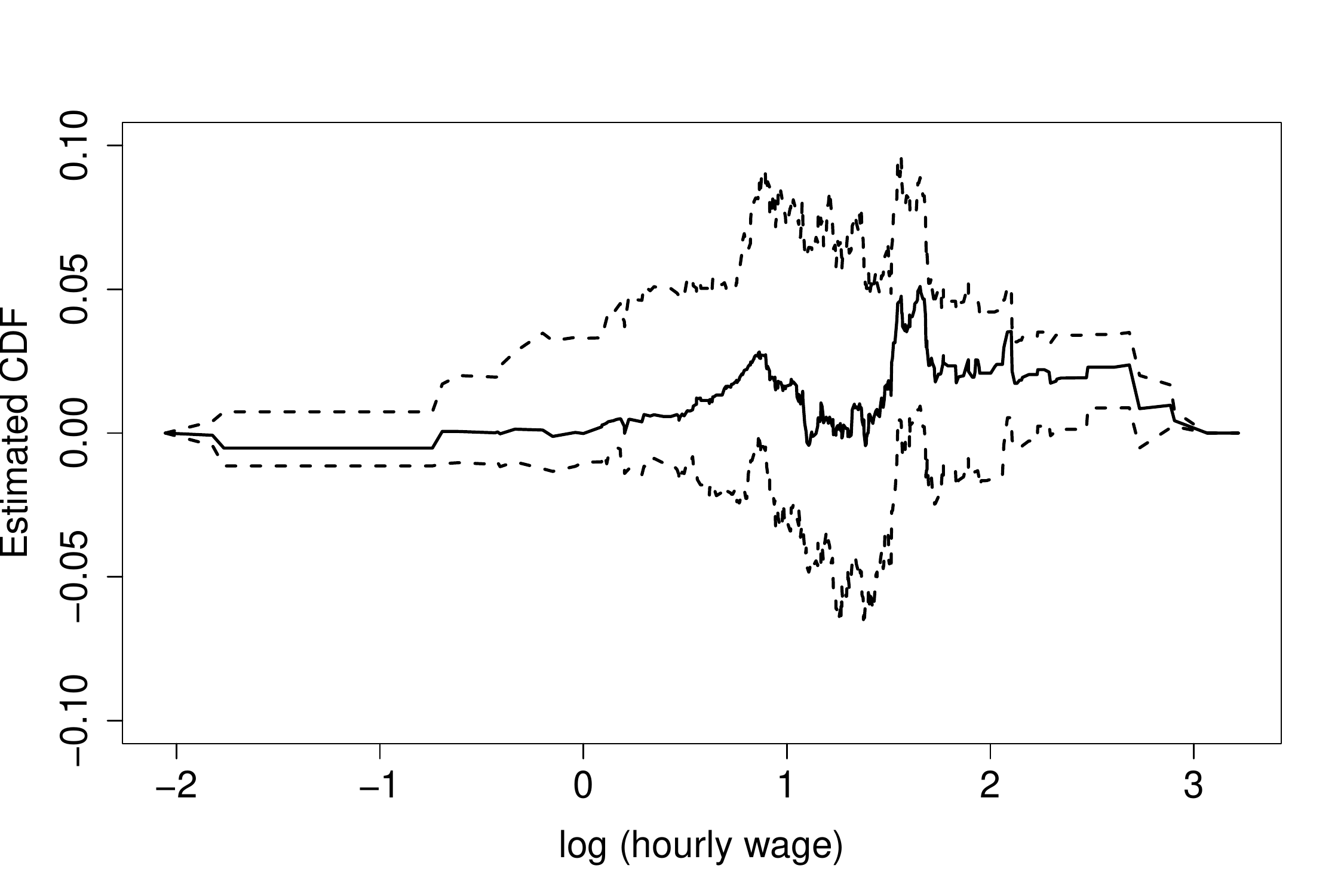}}\\
		\subfloat[$0.9$-quantile of the regressors]{\includegraphics[width = .5\textwidth]{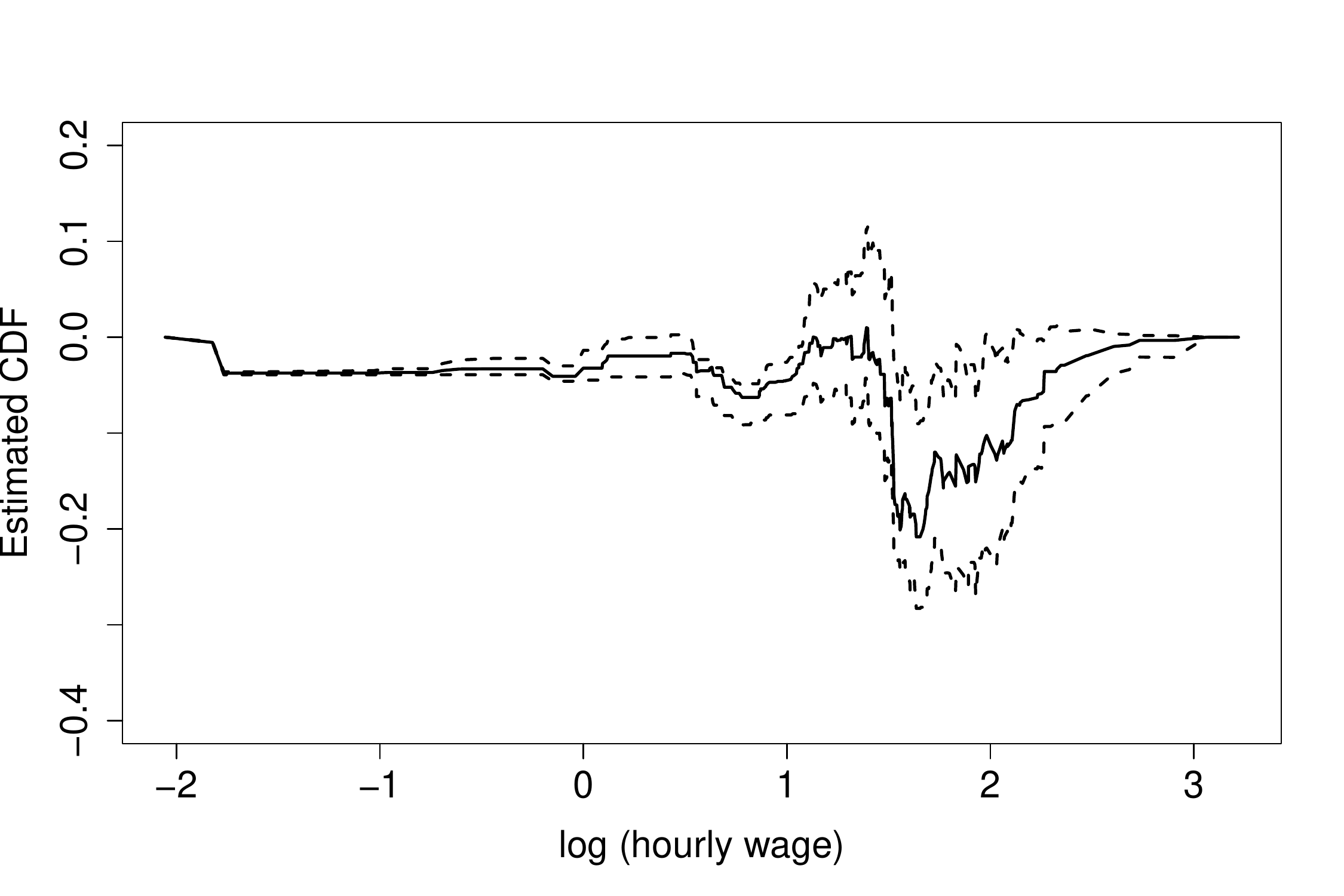}}
    \caption{Difference between estimated conditional distribution functions and confidence bounds}\label{figure:app2}
\end{figure}

To give more evidence about the difference between OLS and IV estimation, pointwise confidence bounds for the differences of the conditional distribution functions are calculated and plotted in Figure \ref{figure:app2}. This is done by bootstrap, i.e. by drawing with replacement $B=200$ times from the individuals. For each $y$, the confidence interval to the level of significance $90\%$ is calculated. This yields a Hausman-type statistical test for the relevance of the IV approach: If $0$ is not contained in the interval, one can conclude that the two estimators of the distribution functions are statistically significantly different. Assuming that the instrument is exogenous and correlated with the endogenous regressor, the IV-based estimator is then the only valid one.

\begin{center}
- Figure \ref{figure:app2} here -
\end{center}

The confidence bounds essentially confirm the analysis from Figure \ref{figure:app1}. For the $10\%$ as well as for the $90\%$ quantile, the bounds do not contain $0$ for some subsets of the ranges of $log(wage)$ described above. For the $50\%$ quantile, the confidence bounds are smaller than these for the $10\%$ and $90\%$ quantile, which is the expected behavior from the simulations and from Table \ref{table:app}). The value $0$ lies outside the bounds for $y$ slightly smaller than $1$. Summed up, the message is that IV-estimation of DR models does make a difference compared to OLS-estimation.

\section{Conclusion and Outlook}
The paper has proposed a new consistent estimator for the semiparametric DR model which allows for endogenous regressors and where monotonicity is enforced. The method is easy to implement and should be appealing to practitioners. Apparently, the proposed procedure only works well if the instruments are sufficiently strong. To circumvent the problem of choosing appropriate instruments, it might be an idea for future research to adapt the procedure proposed by \citet{breitung:2022} for linear regression models to DR models. Here, rank-based transformations of non-normal regressors are used as additional regressors and no external instruments are necessary to obtain consistent parameter estimators. Other tasks for future research would be analytical results for the isotonic regression and a framework for discrete endogenous variables. The latter could be done along the lines of \citet{wooldridge:2002}, Section 15.7.3, but would be computationally harder because there would no simple more step procedure available.

\appendix
\section{Appendix}
The assumptions required for Theorem 1 are formulated in terms of the expected score function $\Psi: \Theta \times \mathcal{I} \rightarrow \Theta$ with $\Theta$ being a compact subset of $\left(\mathbb{R}^{k+l+2} \times (-1,1) \times (0,\infty)\right)$ and an open interval $\mathcal{I}$ that covers a compact interval $\mathcal{U}$. It holds $\Psi(\theta(y),y) = E(\Psi_i^*(\theta(y),y))$ with $\Psi_i^*(\theta(y),y) = \frac{\partial}{\partial \theta(y)} L_{y,i}(\theta(y)) = $
\begin{eqnarray*}
\begin{pmatrix} \frac{X_i}{\sigma_\epsilon} A_{y,i}  \\ \frac{Y_{2,i}}{\sigma_\epsilon} A_{y,i} \\ \frac{X_i}{\sigma_\epsilon} \left(-\rho A_{y,i} + \frac{1}{\sigma_2^2} \left(Y_{2,i} - X_i'\gamma_1 - Z_i'\gamma_2\right)\right) \\ \frac{Z_i}{\sigma_\epsilon} \left(-\rho A_{y,i} + \frac{1}{\sigma_2^2} \left(Y_{2,i} - X_i'\gamma_1(y) - Z_i'\gamma_2(y)\right)\right) \\ \frac{\partial}{\partial \rho}\frac{\mu_{y,i}(\theta(y))}{\sigma_\epsilon}  A_{y,i} \\ \frac{\partial}{\partial \sigma_2^2}\frac{\mu_{y,i}(\theta(y))}{\sigma_\epsilon}  A_{y,i} - \frac{1}{\sigma_2^2} + \frac{1}{2 \sigma_2^4} \left(Y_{2,i} - X_i'\gamma_1(y) - Z_i'\gamma_2(y)\right)^2 \end{pmatrix}
\end{eqnarray*}
and 
\begin{equation*}
A_{y,i} = I_{y,i} \left(\Phi\left(\frac{\mu_{y,i}(\theta(y))}{\sigma_\epsilon}\right)\right)^{-1} \varphi\left(\frac{\mu_{y,i}(\theta(y))}{\sigma_\epsilon} \right) - (1-I_{y,i}) \left(\Phi\left(1-\frac{\mu_{y,i}(\theta(y))}{\sigma_\epsilon}\right)\right)^{-1} \varphi\left(1-\frac{\mu_{y,i}(\theta(y))}{\sigma_\epsilon} \right)
\end{equation*}
\begin{eqnarray*}
\frac{\partial}{\partial \rho}\frac{\mu_{y,i}(\theta(y))}{\sigma_\epsilon} &=& \frac{\sigma_\epsilon \left(Y_{2,i} - X_i'\gamma_1(y) - Z_i'\gamma_2(y)\right) + \left(X_i'\beta_1(y) + Y_{2i} \beta_2(y) + \rho \left(Y_{2,i} - X_i'\gamma_1 - Z_i'\gamma_2\right)\right) \frac{\rho \sigma_2^2}{\sigma_\epsilon}}{\sigma_\epsilon^2} \\
\frac{\partial}{\partial \sigma_2^2}\frac{\mu_{y,i}(\theta(y))}{\sigma_\epsilon} &=& \frac{\left(X_i'\beta_1(y) + Y_{2i} \beta_2(y) + \rho \left(Y_{2,i} - X_i'\gamma_1 - Z_i'\gamma_2\right)\right) \frac{\rho \sigma_2^2}{\sigma_\epsilon}}{\sigma_\epsilon^2}
\end{eqnarray*}
In the remaining part of this appendix, we denote the true parameter by $\theta_0(y)$. We impose similarly to Lemma E.1 and Lemma E.3 in \citet{chernozhukov:2013}
\begin{assumption}
\begin{itemize}
\item[1.] (a) $\Psi: \Theta \times \mathcal{I} \mapsto \Theta$ is continuous, and $\theta \mapsto \Psi(\theta, u)$ is the gradient of a convex function in $\theta$ for each $u \in \mathcal{U}$, (b) for each $u \in \mathcal{U}$, $\Psi\left(\theta_0(u), u\right)=0$, (c) $\frac{\partial}{\partial\left(\theta^{\prime}, u\right)} \Psi(\theta, u)$ exists at $\left(\theta_0(u), u\right)$ and is continuous at $\left(\theta_0(u), u\right)$ for each $u \in$ $\mathcal{U}$, and $\dot{\Psi}_{\theta_0(u), u}:=\left.\frac{\partial}{\partial \theta^{\prime}} \Psi(\theta, u)\right|_{\theta_0(u)}$ obeys $\inf _{u \in \mathcal{U}} \inf _{\|h\|=1}\left\|\dot{\Psi}_{\theta_0(u), u} h\right\|>c_0>0$.
\item[2.] The function class $\mathcal{D} := \{\Psi_i^*(\theta,y): \theta \in \Theta, y \in \mathcal{I}\}$ is Donsker with square integrable envelope.
\end{itemize}
\end{assumption}
Ass.1.1 is required for the pointwise convergence of $\hat \theta(y)$ to $\theta_0(y)$. A crucial point here is the positive definiteness of the derivative matrix of the score vector in Ass. 1.1.(a), from which the existence of a unique Z-estimator follows. Results from \citet{newey:1987} (Assumption A.3.(v)) or \citet{amemiya:1978}, Section 6, yield that this holds in the just identified case if $E_{XZ} := E((X_i, Z_i)(X_i, Z_i)')$ is invertible and if the true parameters lie inside of the parameter space, see \citet{riversvuong:1988}. The invertibility condition implies that weak instruments might be problematic for the estimation procedure.

Ass.1.2 concerns the uniform convergence of $\hat \theta(y)$ to $\theta_0(y)$. Due to the boundedness of $\Theta$, the (then assumed to be finite) norm of the matrix $E_{XZ}$ can be chosen as the envelope in Ass. 1.2., see Step 3 in the proof of Theorem 5.2 in \citet{chernozhukov:2013} and Example 19.7 in \citet{vandervaart:1998}. Then the Donsker property holds with the observation that the function class $\mathcal{D}$ is a Lipschitz transformation of VC classes.\\

\textit{Proof of Theorem 1} \\
Denote $l^{\infty}(\mathcal{A})^p$ the space of $p$-dimensional bounded functions with index sex $\mathcal{A}$. With Assumption 1.2, $\sqrt{n}(\widehat{\Psi}(\cdot,\cdot)-\Psi(\cdot,\cdot)) \Rightarrow_d Z$ in $l^{\infty}(\Theta \times \mathcal{U})^{k+l+4}$, where $Z$ is a Gaussian process. With Assumption 1.1, Condition $Z$ in \citet{chernozhukov:2013} holds and $u \mapsto \theta_0(u)$ is continuously differentiable. Then, with Lemma E.3 in \citet{chernozhukov:2013}, $$
\sqrt{n}\left(\widehat{\theta}(\cdot)-\theta_0(\cdot)\right)=-\dot{\Psi}_{\theta_0(\cdot), \cdot}^{-1} \sqrt{n}(\widehat{\Psi}-\Psi)\left(\theta_0(\cdot), \cdot\right)+o_{\mathbb{P}}(1) \rightsquigarrow-\dot{\Psi}_{\theta_0(\cdot), \cdot}^{-1}\left[Z\left(\theta_0(\cdot), \cdot\right)\right] =: G(\cdot)$$
in $l^{\infty}(\mathcal{U})^{k+l+4}$.

Note that $F_{Y|X,Y_2}(y|x,y_2) := \Phi(x'\beta_{1,0}(y) + y_2 \beta_{2,0}(y))$ is a differentiable transformation of the first $k+1$ components of the vector $\theta_0(y)$. So, we can apply Theorem A.1 in \citet{wied:2012} to express the limit process of \eqref{eq:theorem1} by means of the functional delta method and the gradient of this function. To be precise, we have $f(x_1,x_2)=\Phi(X'x_1 + Y_2 x_2)$ and $D f(x_1,x_2) = \begin{pmatrix} x \varphi(x'x_1 + y_2 x_2) \\ Y_2 \varphi(x'x_1 + y_2 x_2) \end{pmatrix}$. This leads to

\begin{equation}
\sqrt{n}\left(\hat F_{ML,Y|X,Y_2}(\cdot|x,y_2)- F_{Y|X,Y_2}(\cdot|x,y_2) \right) \Rightarrow_d D f(\beta_{1,0}(\cdot),\beta_{2,0}(\cdot))' \begin{pmatrix} G_{1,\ldots,k}(\cdot) \\ G_{k+1}(\cdot) \end{pmatrix} \text{ in } l^{\infty}(\mathcal{U}).
\end{equation}

\hfill $\Box$ \\

\textit{Proof of Theorem 2}\\
We have the integral representation $$\hat F_{Y|X,Y_2}(y|x,y_2) = \int \Phi\left(x'\widehat{\tilde{\beta}_1(y)} + y_2 \widehat{\tilde{\beta}_2(y)} + v \widehat{\tilde{\rho}}\right) d \hat F_n(v),$$ where $\hat F_n(v)$ is the empirical distribution function of the OLS residuals $\hat v_i$. As $V \sim \mathcal{N}(0,\sigma_2^2)$ with distribution function $\Phi_{0,\sigma_2^2}$, the population analogon is $$F_{Y|X,Y_2}(y|x,y_2) = \int \Phi\left(x'\tilde{\beta}_1(y) + y_2 \tilde{\beta}_2(y) + v \tilde{\rho}\right) d \Phi_{0,\sigma_2^2}(v).$$ The empirical process $\sqrt{n}\left(\hat F_n(\cdot) - \Phi\left(\frac{\cdot}{\sigma_2}\right) \right)$ converges to a Gaussian process (see e.g. \citealp{chen:2001}) in $l^{\infty}(\mathbb{R})$. Then the result follows from Theorem 1 and the functional delta method. \hfill $\Box$

\bibliography{bibl}

\end{document}